\documentclass[twocolumn]{aastex631}

\usepackage{enumitem}

\newcommand{\Omegab}{\Omega_\mathrm{b}}

\shorttitle{Globular clusters in bar resonances}
\shortauthors{A. M. Dillamore et al.}

\begin{document}

\title{Trojan globular clusters: radial migration via trapping in bar resonances}

\correspondingauthor{Adam M. Dillamore}
\email{amd206@cam.ac.uk}

\author{Adam M. Dillamore}
\affiliation{Institute of Astronomy, University of Cambridge, Madingley Road, Cambridge CB3 0HA, UK}

\author{Stephanie Monty}
\affiliation{Institute of Astronomy, University of Cambridge, Madingley Road, Cambridge CB3 0HA, UK}

\author{Vasily Belokurov}
\affiliation{Institute of Astronomy, University of Cambridge, Madingley Road, Cambridge CB3 0HA, UK}

\author{N. Wyn Evans}
\affiliation{Institute of Astronomy, University of Cambridge, Madingley Road, Cambridge CB3 0HA, UK}


\begin{abstract}

We search for globular clusters (GCs) trapped in resonances with the bar of the Milky Way. By integrating their orbits in a potential with a decelerating bar, we select 10 whose orbits are significantly changed by its presence. Most of these are trapped in the corotation resonance (CR), including M22 and 47 Tuc. The decelerating bar is capable of transporting these GCs to their current positions from much lower energies, angular momenta, and radii. Our results indicate that the bar is likely to have reshaped the Milky Way's globular cluster system via its resonances. We also discuss implications for the origins of specific GCs, including the possible nuclear star cluster M22. Finally, we consider the effects of the bar on the tidal tails of a trapped GC, by running simulations of stars stripped from 47 Tuc. Instead of forming narrow tails, the stripped stars make up a diffuse extended halo around the cluster, consistent with observations of 47 Tuc.

\end{abstract}



\section{Introduction}
Globular clusters (GCs) are some of the oldest constituents of the Milky Way and other galaxies, and are therefore very useful tools for understanding the early stages of galaxy formation. It has long been known that the GCs can be roughly divided into two groups: those born inside the Milky Way (\textit{in situ}) and those accreted during mergers with satellites. The former have generally been associated with higher metallicities and a higher fraction of disc-like orbits \citep{zinn1985}. More recent studies have used a combination of kinematics, the age-metallicity plane, and chemical abundances to divide the population of GCs into these two groups \citep{dinescu1999,marinfranch2009,forbes2010,leaman2013,recioblanco2018,massari2019, Cal22}. Informed by aluminium-to-iron ratios [Al/Fe], \citet{belokurov2023,belokurov2024} separated the GCs into \textit{in situ} and accreted components in the space of energy $E$ vs angular momentum $L_z$, with \textit{in situ} GCs at lower energies. If the Galactic potential is axisymmetric, $E$ and $L_z$ are integrals of motion, and remain conserved for each cluster.

However, the Milky Way's potential is not axisymmetric. The centre of the Galaxy hosts a rotating bar, which causes $E$ and $L_z$ to vary with time. These effects are particularly important if an orbit is in \textit{resonance} with the bar. A set of resonances with a bar rotating at a pattern speed (angular frequency) $\Omegab$ can be defined by
\begin{equation}\label{eq:resonances}
    m(\Omega_\phi-\Omegab)+l\,\Omega_r=0.
\end{equation}
$\Omega_\phi$ and $\Omega_r$ are the azimuthal orbital frequency and radial oscillation frequency respectively, and $l$ and $m$ are integers. Important resonances include the corotation resonance (CR; $l=0$, $\Omega_\phi=\Omegab$) and the inner and outer Lindblad resonances (ILR and OLR; $l/m=\mp1/2$). Stable orbits near the CR are usually centred on the $L_4$ and $L_5$ Lagrange points \citep[see pp.178-184 in][]{binney_tremaine}. Particles can become trapped by these resonances, creating overdense structures in phase space \citep{De00,sellwood2010,Ge11,Mc13,binney2020,chiba2021,chiba2021_treering}. Much attention has been given to such structures in the stellar disc, particularly since data from the \textit{Gaia} observatory \citep{gaia} became available \citep{Fr19,khoperskov2020,trick2021,Tr22,Kh22,wheeler2022}. However, stars orbiting in the halo can also become trapped, creating moving groups and substructure on more eccentric and inclined orbits \citep{moreno2015,moreno2021,dillamore2023,dillamore2024}. It is therefore natural to consider whether any of the Milky Way's globular clusters are trapped in bar resonances. While \citet{bajkova2023} and \citet{smirnov2024} have considered the dynamics of GCs located in the bar itself, its influence extends well beyond the Galaxy's inner regions and may affect clusters at larger radii. \citet{tkachenko2023} also studied the effects of the bar on GC orbits, but did not consider resonances or a decelerating bar. By analogy to trojan asteroids orbiting $L_4$ and $L_5$ Lagrange points in the Solar System, we refer to any GCs trapped at the CR as \textit{trojan globular clusters}, although we also consider trapping at other resonances.

A rotating bar is expected to decelerate with time due to dynamical friction with the dark matter halo \citep{Deb98,athanassoula2002,We02,Ce07,collier2019,chiba2022,hamilton2023} A bar with a decelerating pattern speed can drag particles trapped in its resonances, transporting them to higher energies and angular momenta \citep{chiba2021,chiba2021_treering}. The Galactic bar is therefore endowed with the ability to reshape the distribution of particles in integral-of-motion space. This challenges the notion that the current positions of GCs in $(L_z,E)$ space can be used to directly infer their origin and, by extension, the assembly history of the Milky Way \citep[e.g.][]{MySausageGC,massari2019,belokurov2023,belokurov2024}. Our aim is to determine which GCs are most likely to have been affected by the bar, and how their orbits may have been changed by a decelerating pattern speed. This could affect our understanding of the origin of the Milky Way's globular cluster system.

Tidal tails (or stellar streams) are formed when stars escape from a GC through its Lagrange points, forming one or two extended structures approximately aligned with the GC's orbit. This makes them extremely valuable tools for studying the Galaxy. They have been used for measuring the Milky Way's potential \citep[e.g.][]{johnston1999,koposov2010,gibbons2014,bowden2015,bonaca2018} and to probe the distribution of dark matter \citep[e.g.][]{ibata2002,erkal2015,erkal2016,bovy2016,Bonaca_GD1}. The Milky Way's bar may also perturb some tidal tails, creating gaps, asymmetry, and fanning via chaotic effects \citep{Ha16,pricewhelan2016,pearson2017}. It is therefore worth considering the effects of the bar on tidal tails formed from GCs trapped in resonances.

The rest of this \textit{Letter} is arranged as follows. In Section~\ref{section:data_sim} we describe the data and test particle simulation used in this study. We integrate the orbits of the GCs in a decelerating barred potential and analyse the results in Section~\ref{section:orbits}. Individual GCs of interest are discussed in Section~\ref{section:individual}, including their possible origins. In Section~\ref{section:stream} we run simulations of tidal tails released from a cluster trapped in a resonance. Finally, we summarise our conclusions in Section~\ref{section:summary}.

\begin{figure}
  \includegraphics[width=\columnwidth]{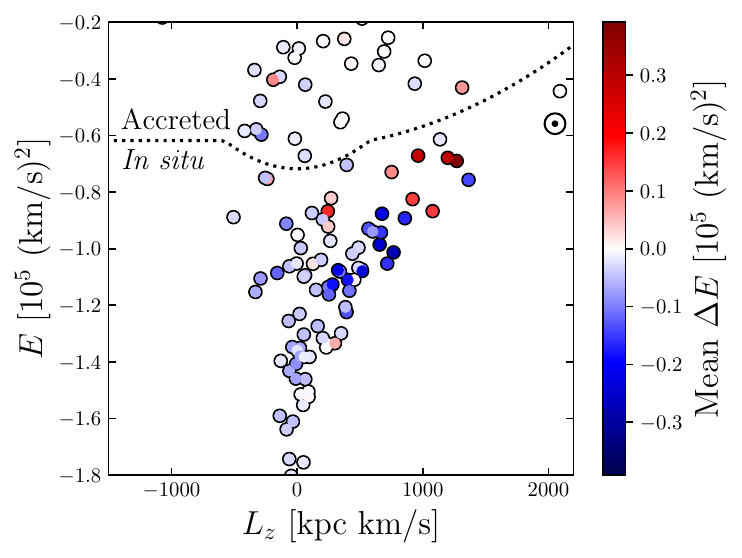}
  \caption{Energy $E$ vs angular momenta $L_z$ of GCs in our sample. The black dotted line marks the boundary between \textit{in situ} (lower $E$) and accreted (higher $E$) clusters. The colours indicate mean change in energy $\Delta E$ as the bar's pattern speed slows from $\Omegab=80$ to 35~km\,s$^{-1}$\,kpc$^{-1}$. Most of the GCs with $\Delta E>0$ are located around $L_z\sim1000$~kpc\,km\,s$^{-1}$ and $E\sim-0.7\times10^5$~km$^2$\,s$^{-2}$.}
  \label{fig:Delta_E_all}
\end{figure}

\begin{figure*}
  \centering
  \includegraphics[width=\textwidth]{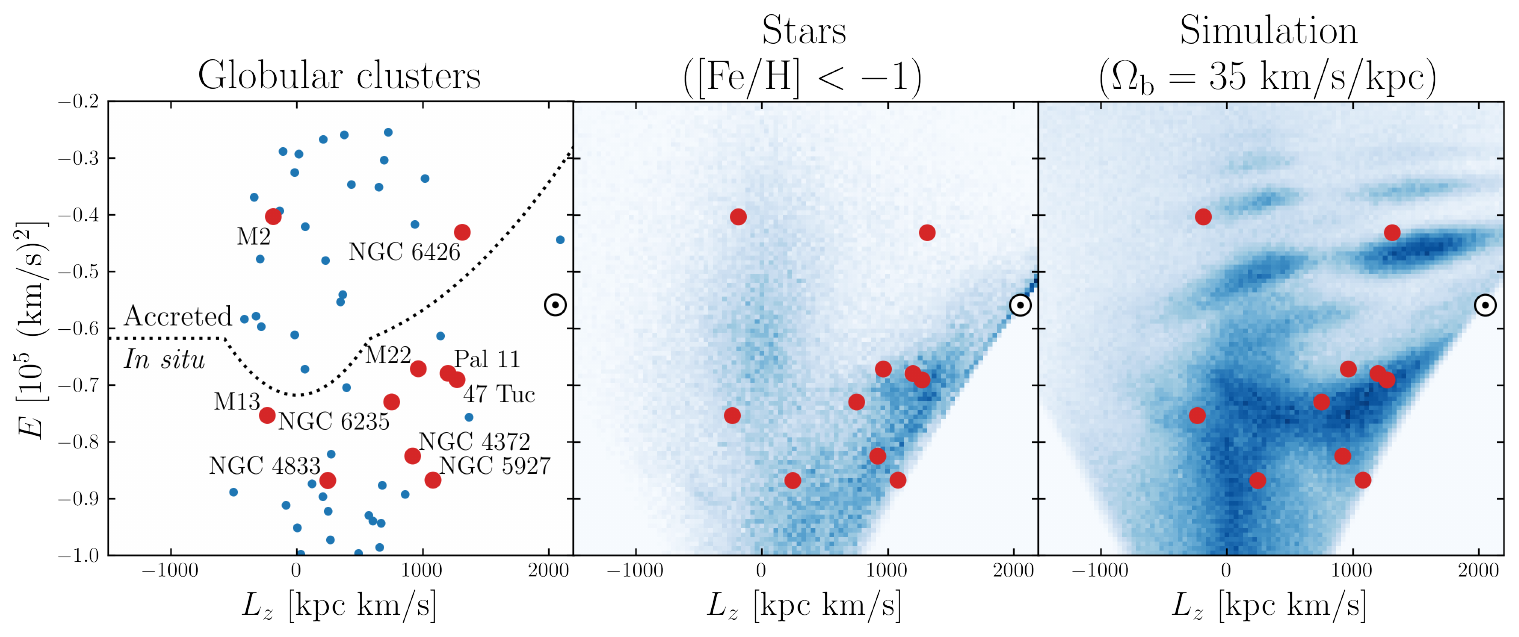}
  \caption{Present day energy $E$ vs $z$-component of angular momentum $L_z$ for GCs (left-hand panel), stars with [Fe/H]~$<-1$ (middle panel), and our test particle simulation with a rotating bar (right-hand panel). In each panel the large red points mark the 10 selected GCs with the largest mean changes in $L_z$ due to the slowing bar. The black dotted line in the left-hand panel marks the boundary between accreted (higher $E$) and \textit{in situ} (lower $E$) GCs according to \citet{belokurov2023,belokurov2024}, and the Sun is marked with a $\odot$ symbol.}
  \label{fig:E_Lz}
\end{figure*}

\section{Data and simulations}
\label{section:data_sim}
\subsection{Globular clusters}
We use GC proper motions published by \citet{vasiliev2021} derived from \textit{Gaia} EDR3 measurements \citep{gaia_edr3}, with positions from \citet{harris2010}. Distances and line-of-sight velocities are taken from \citet{baumgardt2021} and \citet{baumgardt2019} respectively. This provides us with 6D coordinates and associated uncertainties for 170 objects identified as GCs. We restrict our sample to GCs whose orbits are well-constrained  by including only those whose fractional uncertainty in heliocentric distance $D$ is less than $5\%$ ($\sigma_D/D<0.05$). This limits the sample to 141 clusters.

Following \citet{Po17}, we place the Sun in the Galactic plane at a distance $R_0=8.2$~kpc from the Galactic centre, at an angle of $30^\circ$ relative to the bar's major axis \citep[consistent with][]{wegg2015}. The circular velocity at the Sun's radius is $v_\mathrm{c}(R_0)=238$~km\,s$^{-1}$ \citep{bland-hawthorn2016}, and the Sun's velocity relative to the local standard of rest is $(U,V,W)_\odot=(11.1,12.24,7.25)$~km\,s$^{-1}$ \citep{schonrich2010}. We work in a left-handed Galactocentric coordinate system in which the disc has $L_z>0$.

Throughout this paper we use the the analytic Milky Way potential fitted by \citet{sormani2022} to the $N$-body barred model of \citet{Po17}. This potential includes multiple bar components, a disc, a central mass concentration (representing a nuclear stellar disc or cluster), and a flattened axisymmetric dark matter halo. We calculate all energies $E$ and angular momenta $L_z$ in the axisymmetrised (i.e. azimuthally averaged) version of this potential. The $(L_z,E)$ values of the GCs in our sample are shown in Fig.~\ref{fig:Delta_E_all} and in the left-hand panel of Fig.~\ref{fig:E_Lz} (discussed in more detail in Section~\ref{section:orbits}). The black dotted lines mark the approximate boundary between \textit{in situ} and accreted GCs in this potential \citep{belokurov2023,belokurov2024}.

\subsection{Stars}
We compare the distribution of globular clusters in phase space with that of stars in the Milky Way. We select stars from the third data release of \textit{Gaia} \citep[DR3;][]{gaia,gaia_dr3} with line-of-sight velocity measurements from the Radial Velocity Spectrometer \cite[RVS;][]{gaia_rvs}. We use estimates of distance $D$, metallicity [Fe/H] and surface gravity $\mathrm{log}\,g$ derived from XP spectra by \citet{zhang2023}. We select sources with line-of-sight velocity measurements, distance uncertainties $<10\%$, distances $D<15$~kpc, and \texttt{quality\_flags}~$<8$ as recommended by \citet{zhang2023}. Sources within $1.5^\circ$ of known GCs less than $5$~kpc from the Sun are also removed. To subtract the thin disc from the sample and reveal halo substructure we focus on stars with [Fe/H]~$<-1$. In addition, we include only giant stars with $\mathrm{log}\,g<3$, in order to remove the strong spatial selection effect around the Sun. The distribution of the sample in $(L_z,E)$ space is shown in the middle panel of Fig.~\ref{fig:E_Lz} in blue.

\subsection{Test particle simulation}
The distributions of GCs and stars in $(L_z,E)$ space are also compared to a test particle simulation of the halo and thick disc in the presence of a slowing rotating bar. This simulation is described in detail in \citet{dillamore2024} and briefly summarised below.

We initialise a steady-state distribution of stars in the axisymmetrised \citet{sormani2022} potential described above. This distribution consists of a non-rotating halo-like component and a rotating thick disc-like component. We integrate the orbits in this potential with the addition of a bar which smoothly increases in strength over $\approx2$~Gyr, then smoothly decelerates in pattern speed $\Omegab$. This bar consists of the non-axisymmetric multipole components of the \citet{sormani2022} potential, with their spatial scale inversely proportional to $\Omegab$ such that the bar length increases with time. Over the course of the simulation, the pattern speed decreases from $\Omegab=80$~km\,s$^{-1}$\,kpc$^{-1}$ to $35$~km\,s$^{-1}$\,kpc$^{-1}$. See \citet{dillamore2024} for full details.

The $(L_z,E)$ distribution of the final snapshot of the simulation (when $\Omegab=35$~km\,s$^{-1}$\,kpc$^{-1}$) is shown in blue in the right-hand panel of Fig.~\ref{fig:E_Lz}. As discussed by \citet{dillamore2024}, the roughly horizontal ridges correspond to resonances with the bar, including the corotation resonance (CR; $E\approx-0.7\times10^5$~km$^2$\,s$^{-2}$) and the outer Lindblad resonance (OLR, $E\approx-0.5\times10^5$~km$^2$\,s$^{-2}$).

\begin{figure*}
  \centering
  \includegraphics[width=\textwidth]{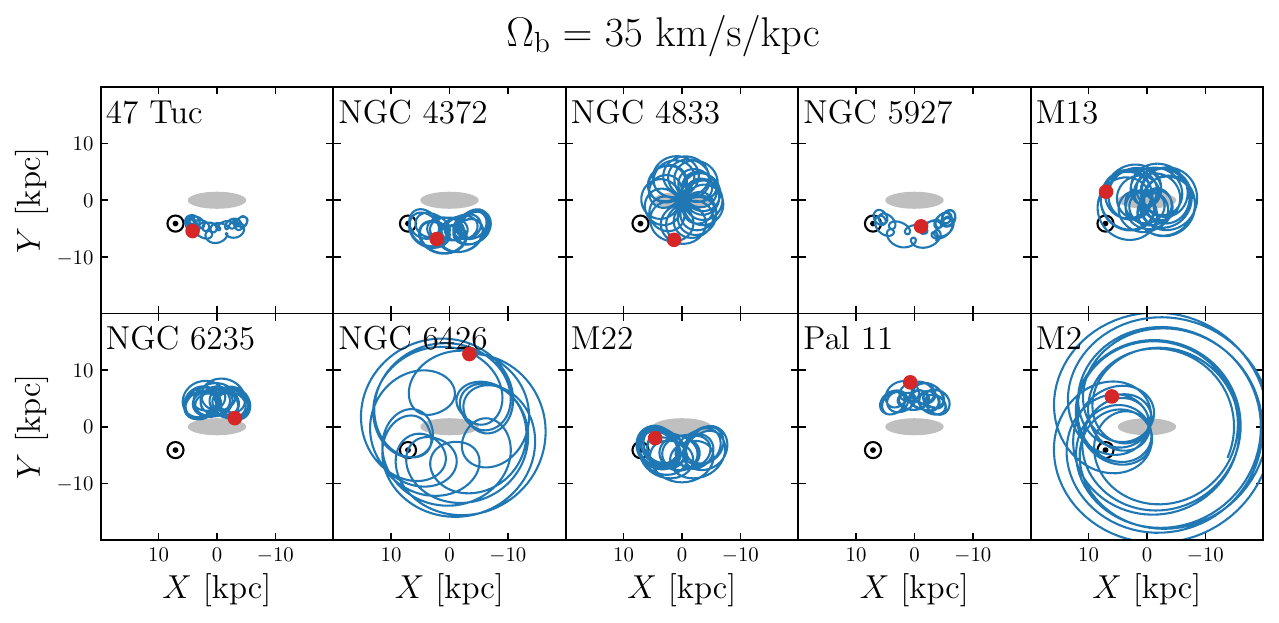}
  \caption{Orbits of the 10 selected GCs in the \citet{sormani2022} potential rotating with constant pattern speed $\Omegab=35$~km\,s$^{-1}$\,kpc$^{-1}$. The frame shown is corotating with the bar (grey ellipse), and the current positions of the Sun and the GCs are shown with the $\odot$ and red points respectively. Six of the 10 GCs (e.g. 47 Tuc) orbit close to the corotation resonance (CR), since they remain on one side of the bar.}
  \label{fig:orbits}
\end{figure*}

\begin{figure*}
  \centering
  \includegraphics[width=\textwidth]{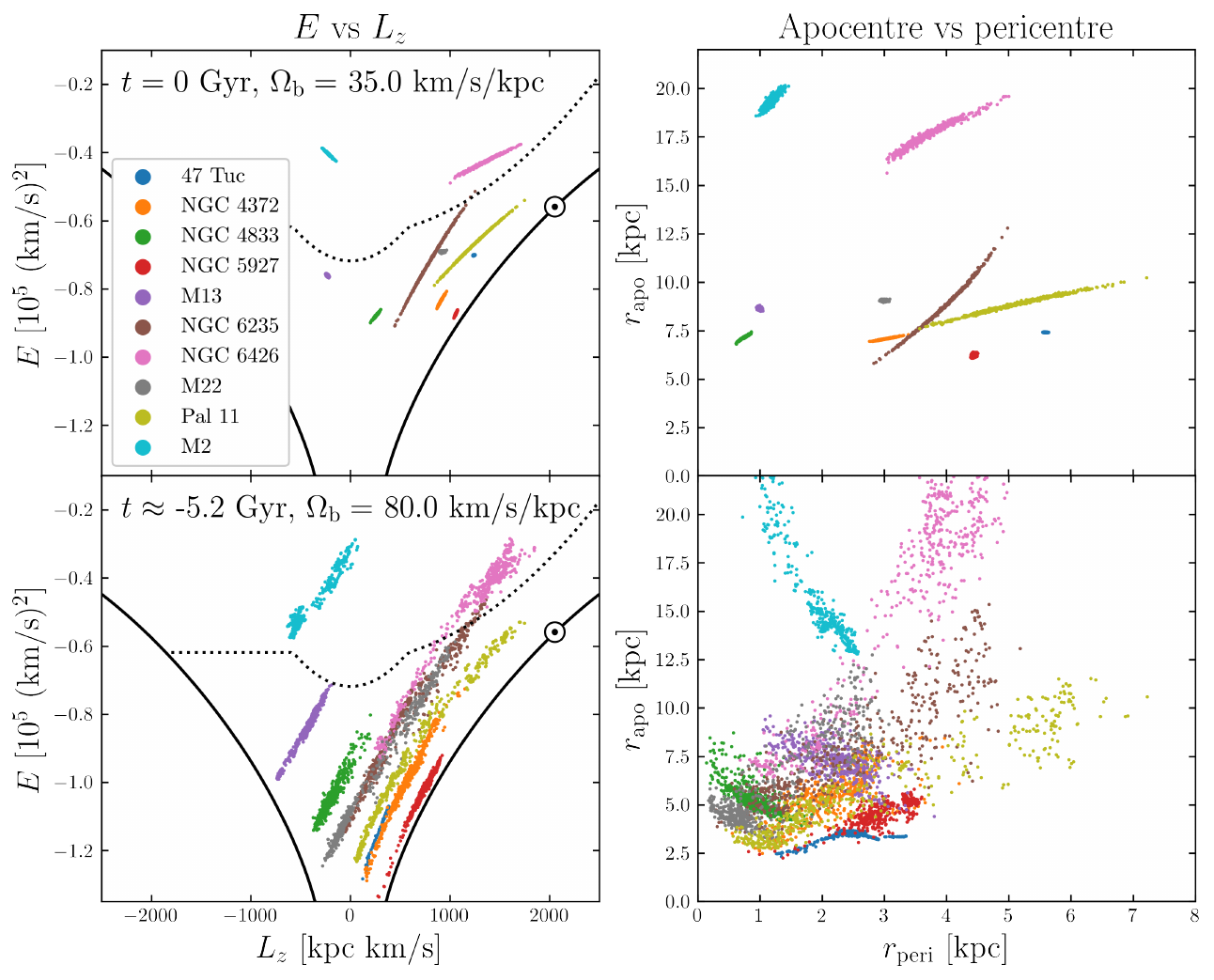}
  \caption{Changes in the orbits of the 10 GCs in a slowing barred potential. The top and bottom rows show the present day $t=0$ and $t=t_0\approx-5.24$~Gyr, at which the pattern speeds are $\Omegab=35$ and 80~km\,s$^{-1}$\,kpc$^{-1}$ respectively. The left and right-hand columns show energy vs angular momentum and apocentre vs pericentre respectively. Each of the 10 clusters is on average at lower energy and angular momentum at the earlier time. In some cases (e.g. M22) the pericentres are also considerably smaller than their current values.}
  \label{fig:orbit_changes}
\end{figure*}

\section{GC orbits in a barred potential}
\label{section:orbits}
Our aim is to find GCs in our sample whose orbits have a high probability of being trapped in resonances with the bar. In particular, we are interested in those which may have been transported from smaller radii and lower energies by a decelerating bar. We achieve this by integrating their orbits backwards in time in a potential with a slowing bar (i.e. decreasing in pattern speed from the past to the present day). Our method is described below.

\subsection{Potential with a slowing bar}
\label{section:slowing_potential}
We again use a modified version of the \citet{sormani2022} barred potential to integrate the GC orbits. We characterise the slowing of the pattern speed $\Omegab$ with the dimensionless deceleration parameter, $\eta\equiv-\dot{\Omega}_\mathrm{b}/\Omegab^2$. This is held at $\eta=0.003$ at all times, consistent with \citet{chiba2021}. The scale-length of the bar $S$ varies in time according to $S\propto1/\Omegab$, and matches the \citet{sormani2022} model when $\Omegab=39$~km\,s$^{-1}$\,kpc$^{-1}$. This dependence is similar to previous studies of a slowing bar, e.g. \citet{chiba2021}. The amplitude of the bar's potential is kept constant throughout. This potential is identical to that used for the simulations, except that we keep the amplitude and deceleration parameter $\eta$ constant throughout, instead of smoothly changing them from zero at early times. The pattern speed and strength evolution is otherwise identical, so these differences will not significantly affect the comparison with the simulation at the present day.

The orbit integrations are started at the present day $(t=0)$ with a pattern speed of $\Omegab=35$~km\,s$^{-1}$\,kpc$^{-1}$, consistent with our test particle simulation and various recent studies of the Milky Way's bar \citep{Sa19,binney2020,chiba2021_treering}. We integrate the orbits backwards in time to $t=t_0\approx-5.24$~Gyr, when $\Omegab=80$~km\,s$^{-1}$\,kpc$^{-1}$. This is likely to be less than the age of the bar \citep{sanders2023}, so our estimates for the changes in the orbits are likely to be conservative. We note that the past evolution of the pattern speed remains poorly constrained, so quantities evaluated at this earlier snapshot should be viewed as qualitative results. We sample from the uncertainties in the 6D phase space measurements of the GCs \citep[provided by][]{baumgardt2019,baumgardt2021,vasiliev2021}, assuming each measurement is uncorrelated and Gaussian distributed with appropriate variance. We integrate 500 realizations for each cluster.

\subsection{Results}
For each cluster in our sample we calculate the energies and angular momenta of each realization when $\Omegab=35$~km\,s$^{-1}$\,kpc$^{-1}$ and 80~km\,s$^{-1}$\,kpc$^{-1}$. We use the change in energy $\Delta E$ between $t=t_0$ and $t=0$ as a measure of the change in each orbit. We compute the mean $\Delta E$ across all realizations of each cluster. These values are shown as a function of $L_z$ and $E$ for GCs in our cut in Fig.~\ref{fig:Delta_E_all}. As the bar slows, particles trapped in most resonances move to larger radii, $E$ and $L_z$. In this study we are interested in the GCs with $\Delta E>0$, most of which lie around $L_z\sim1000$~kpc\,km\,s$^{-1}$ and $E\sim-0.7\times10^5$~km$^2$\,s$^{-2}$. To study these further we select the 10 GCs with the largest positive mean $\Delta E$. This set includes some of the largest and brightest GCs, such as 47 Tucanae (47 Tuc, NGC 104), M2 (NGC 7089), M13 (NGC 6205), and M22 (NGC 6656). We note that there are also several clusters at lower energy which experience a decrease in $E$ as the bar slows. This may be due to trapping in the inner Lindblad resonance, in which $L_z$ decreases as the bar slows \citep{chiba2021}. We leave the study of these for a future work.

We show these 10 GCs with large red points in each panel of Fig.~\ref{fig:E_Lz}. Eight are considered to be \textit{in situ} clusters by \citet{belokurov2024}. Four of these (47 Tuc, M22, NGC 6235 and Pal 11) are located in an overdensity of metal-poor stars (middle panel). Named `Shakti' by \citet{malhan24}, this overdensity has previously been associated with the Hercules stream \citep{My18,dillamore2024}, and is likely due to trapping in a resonance with the bar \citep{myeong2022_eccentric,dillamore2023,dillamore2024}. A similar overdensity at the same position can be seen in the simulation in the right-hand panel, which corresponds to the CR \citep{dillamore2024}. NGC 4372 and 5927 are also located on the lower edge of this overdensity. The other two \textit{in situ} GCs in our selection are at lower $|L_z|$, but are also located in overdense regions of the simulation distribution.

The two accreted clusters are at considerably higher energy and do not overlap with any strong overdensities in the metal-poor stars. However, they each lie on a different resonant overdensity in the simulated distribution. NGC 6426 and M2 lie on the ridges corresponding to the prograde OLR and retrograde 1:2 resonance respectively \citep[see][]{dillamore2024}. The lack of resonant ridges in the data at these high energies may be due to a combination of selection effects and measurement errors, resulting from the greater typical distances of stars at these energies. We note that the positions of these higher energy ridges are sensitive to both the pattern speed and the assumed potential, so the association of these two clusters with these resonances is tentative.

In Fig.~\ref{fig:orbits} we show the orbits of the 10 GCs integrated in the \citet{sormani2022} potential at a constant pattern speed of $\Omegab=35$~km\,s$^{-1}$\,kpc$^{-1}$. The orbits are shown projected onto the Galactic plane in the frame corotating with the bar. Six of these orbits are clearly trapped near the CR, as they remain on one side of the bar instead of circulating around it. The orbit of M2 is also visibly at the retrograde 1:2 resonance, and NGC 6426 is not far from an OLR orbit \citep[c.f. Fig. 6 in][]{dillamore2024}. These are all consistent with the simulation in Fig.~\ref{fig:E_Lz}. The orbits of M13 and NGC 4833 cannot be so easily associated with specific resonances without frequency analysis.

Fig.~\ref{fig:orbit_changes} shows how the orbits of the 10 GCs change between the two snapshots of the slowing bar simulation. The many realizations of each cluster are shown as clouds of coloured points. The top and bottom rows show the present day and the snapshot when $\Omegab=80$~km\,s$^{-1}$\,kpc$^{-1}$ respectively. The left and right columns show ($L_z,E$) and apocentre vs pericentre respectively. As expected, all of our selected clusters are transported to their current positions from lower $E$ and $L_z$. 47 Tuc is transported by the bar in every single realization, while in the other cases there is at least some probability that they were not significantly moved. A few prograde clusters (e.g. M22 and NGC 4833) may have been transported so far that they were originally retrograde ($L_z<0$). The bar may therefore be partially responsible for the current net angular momentum of the globular cluster system. The apocentres and pericentres are also affected. For the prograde clusters these tend to be smaller at the higher pattern speed, due to the lower energy and $L_z$. Some realizations of M22 and NGC 4833 even had pericentres close to zero at the earlier snapshot. For the retrograde clusters (M2 and M13), integrating their trapped orbits back in time decreases $E$ and $L_z$ but increases $|L_z|$, so their orbits become more circular. This results in a larger pericentre but smaller apocentre at the earlier snapshot. We therefore see that the bar is likely to have transported at least some of these clusters from lower energies and radii to their present positions. We note that most of these clusters are classified as \textit{in situ} (see Fig.~\ref{fig:E_Lz}), so transportation from lower energy would not change the interpretation of their \textit{in situ}/accreted origin. A possible exception is NGC 6426, which is classified as accreted but reached below the \textit{in situ} threshold in some realizations.

\section{Discussion of individual GCs}
\label{section:individual}
In our set of bar-interacting GCs, we find a large spread in both metallicity and age. Using metallicities from the 2010 version of the Harris catalogue\footnote{\url{https://physics.mcmaster.ca/~harris/mwgc.dat}} \citep{Ha96}, we find three metallicity groups within our sample. Three of the GCs are metal-poor ($\mathrm{[Fe/H]}\leq-1.7$), four have intermediate metallicities ($-1.7<\mathrm{[Fe/H]}\leq-1.2$) and three display metallicities higher than $\mathrm{[Fe/H]}>-0.7$. Ages, taken from \citet{Va13} for six of our ten GCs, span 10.75 - 12.5~Gyr. Among the GCs without ages in \citet{Va13}, Pal~11 has been suggested to be even younger than NGC~5927 \citep[$t<10$~Gyr, ][]{Le06}, while the most metal-poor GCs in our sample, NGC~4372 and NGC~6426 are likely older than 12.5~Gyr. Additional characteristics and potential implications of bar-interactions on our sample are discussed in the following section. To compare physical characteristics we make use of structural parameters from the fourth version of the Globular Database compiled by H.~Baumgardt\footnote{\url{https://people.smp.uq.edu.au/HolgerBaumgardt/globular/}} \citep{baumgardt2018}. We split the GCs into the three metallicity groups, discussing each in turn. 


\subsection{Metal-Rich Group: Pal~11, NGC~5927, 47~Tuc (NGC~104)}
\label{sec:metalrich}
The three most metal-rich GCs in our sample span the largest range of physical characteristics of each of the three groups. Pal~11 is the least-massive GC in our sample, with a present day mass 60 times smaller than the most-massive GC in the set of ten, 47~Tuc. Interestingly, the two GCs also represent the most (47~Tuc), and least dense (Pal~11) GCs in the set of ten. If we consider 47~Tuc to be the quintessential GC \citep[one of the most massive of the simpler `Type I', nearly chemically homogeneous clusters,][]{Mi2017, Ma2019}, its high central density likely reflects the natural progression of mass segregation within the cluster \citep{Wo56, He71, Ta95}. In other words, the state of the cluster is likely dominated by internal dynamical evolution with little-to-no influence from external forces \citep[e.g. tidal fields,][]{Gn97}. 

Pal~11 represents the counter-example. Given its low density, and relatively short dissolution time \citep[3.4~Gyr, calculated using Equation 10 from][]{baumgardt03}, it likely would not have survived to present-day given the strong tidal forces of the inner Galaxy. The large change in position of Pal~11, with a total gain of $\sim3.2$~kpc in the average radial position between our first and last simulation snapshots, may be responsible for its survival. We suggest that trapping and migration in the corotation resonance may have prevented an earlier dissolution of Pal~11, had it remained in the inner galaxy.

Finally, \citet{Le06}, note that the colour-magnitude diagram (CMD) of NGC~5927, the final GC in our metal-rich group, is remarkably similar to Pal~11. They suggest that this is because the two GCs share very similar metallicities and ages, supported by the metallicities and ages found in the Harris catalogue (2010 edition) and \citet{Va13} respectively. Despite these similarities, the physical characteristics of the two GCs are significantly different. NGC~5927 is very close to the `typical' GC in our set, having a mass and core radius very close to the average values ($\mathrm{M}_{\mathrm{ave}}=3.3\times10^{5}\mathrm{M}_{\odot}$, $r_\mathrm{c}=1.8$~pc). Given the similar ages and metallicities of these two GCs, we suggest that the differences between the two further reinforce the idea that Pal~11 may have experienced a different dynamical history, by way of a larger radial migration. 




\subsection{Intermediate Metallicity Group: NGC~6235, M13 (NGC~6205), M2 (NGC~7089), M22 (NGC~6656)}
Of the intermediate GCs in our sample, two display average global characteristics (core radius, density and mass). These are M13 (NGC~6205) and NGC~6235. Interestingly, our simulations indicate that M13 is the only GC that could have gone from being strongly retrograde to having near-zero $L_{z}$ through interactions with the bar. The two other GCs in this metallicity group represent two of the most exciting GCs found in our set of ten, M22 (NGC~6656) and M2 (NGC~7089). These two GCs have been flagged as potential nuclear star clusters (NSCs), given the presence of metallicity spreads in both. At present, a statistically significant spread in [Fe/H] is the primary feature for NSC candidates in the MW \citep[e.g. in the recent study of][where dynamics were used as a secondary discriminator]{Pf21}, following suggestions made in \citet{Da16}. 

A spread in metallicity has been suggested in M22 for many years \citep{Da09, Ma2011} and was most recently confirmed using high-precision differential abundance analysis \citep[][see their Table~1 for a summary of the history of metallicity spreads in M22]{Mc2022}. \citet{Mc2022} suggest that additional chemical anomalies they find further support M22 being a NSC, possibly of the primordial MW. Furthermore, they also discuss the chemical similarities between M22 and the low-metallicity tail of \textit{Aurora}, the primordial component of the MW occupying the inner halo \citep[$\mathrm{[Fe/H]}\sim-1.7$;][]{belokurov_kravtsov}. Suggestions that the bulk of Aurora stars formed in since-disrupted GCs \citep{belokurov2023} and that the MW was potentially assembled from smaller `building blocks' \citep{Ho24} also support the NSC origin of M22. Finally, given that some of our realizations place the original pericentre values of M22 very close to zero, we support the suggestion of \citet{Mc2022} that M22 is an ancient building block of the MW.

A history of bar interaction for the other NSC candidate in our sample, M2 (NGC~7089), is very interesting given its present day position in $(L_z,E)$ space. Various studies have identified M2 as an accreted GC based primarily on its location in $(L_z,E)$ space \citep{My19, massari2019, Cal22,belokurov2024}, with a strong association with the GSE merger \citep{belokurov2018, helmi2018}. Chemically, M2 shows spreads in both metallicity and a bimodality in heavy, $s$-process elements \citep{Yo14}. Interestingly, M22 also shows a strong bimodality in $s$-process elements \citep{Ma2011, Mc2022}. Although M2 did not originate in the inner MW in our simulations (unlike M22), its chemical complexity and NSC-like properties may also be linked to formation in the primordial MW. Regardless, M2 demonstrates the importance of considering bar interactions when investigating the origin of apparently `accreted' GCs.

\begin{figure}
  \includegraphics[width=\columnwidth]{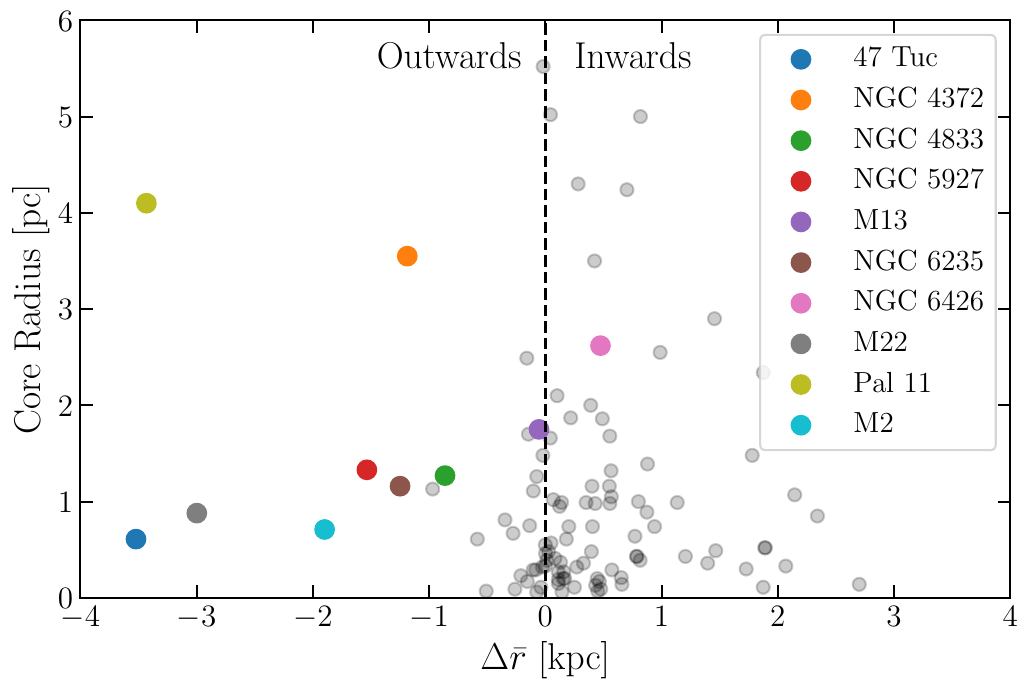}
  \caption{Cluster core radii taken from \citet{baumgardt2018} as a function of the average change in position between our initial and final snapshots (initial - present day, such that outward-moving clusters have $\Delta\bar{r}<0$). Note the possible trend of increasing core radius (reflecting little internal dynamical evolution) as a function of decreasing change in position (change in tidal forces) among our trapped GCs. The only outliers to the trend are the young, metal-rich GC Pal~11 (discussed in Section~\ref{sec:metalrich}), and the ancient, metal-poor GC NGC~4372 (Section~\ref{sec:metalpoor}). Possible explanations for both are given in the corresponding sections.} 
  \label{fig:rcorerelation}
\end{figure}

\subsection{Low Metallicity Group: NGC~4833, NGC~6426, NGC~4372}
\label{sec:metalpoor}
The remaining GCs in our sample have an average metallicity of $\mathrm{[Fe/H]=-2.05}$ \citep[][Harris 2010 edition]{Ha96} and are likely the oldest GCs in our sample \citep[$\sim12.5$~Gyr, ][]{Va13}. Of these three GCs, NGC~4372 stands out from the set given its large core radius and low central density \citep{baumgardt2018}. In fact, with the exception of Pal~11 (which is significantly younger), NGC~4372 is the `fluffiest' (largest core radius) GC in our sample. 

In Fig.~\ref{fig:rcorerelation} we show the distribution of core radii as a function of the change in average orbital radius (initial - present day) for the entire sample of GCs in-common between our simulation and the catalogue of \citet{baumgardt2018}, as well as our trapped GCs. The average orbital radius is here defined as (median apocentre + median pericentre)/2, where the medians are taken across all orbit realizations. A trend of increasing core radius with decreasing change in position is present in the majority of the trapped GCs, while no obvious trend is present in the remainder of the sample. This may suggest that resonance trapping and changing external tidal forces could affect internal cluster evolution, though it is unlikely to be the dominant driver \citep{kremer19}. This trend may also be the result of the birth conditions of the trapped GCs, with the denser GCs having been born closer to the Galactic centre.

The two exceptions to this trend are Pal~11 and NGC~4372. \citet{Ma03}, discuss the appearance of ancient, `fluffy' GCs in their study of GCs around the LMC. They resolve two sequences occupying unique trends in core radius vs. age previously seen by \citet{El91}. One population evolves towards core collapse with increasing age, while the other shows a linear trend of increasing core radius with increasing cluster age. \citet{Ma03} speculated that `normal' cluster evolution (in isolation) represents the first sequence, evolution towards core collapse. The other sequence (to which NGC~4372 belongs) likely represents GCs with varied dynamical histories, different initial mass functions, or large binary fractions. 

Several studies have investigated the prevalence of stellar-mass black holes (BHs) in GCs and their influence on the growth of cluster core radii \citep{merritt04, mackey07, mackey08, breen2013, morscher2015} and velocity dispersion profiles of GCs \citep{Ba23, Di23, Di24}. A `bottom-light' IMF at the time of formation would increase the number of stellar-mass BHs in the cluster. Upon sinking to the GC centre these would increase the central velocity dispersion. In this way, the GC can remain `fluffy'. In fact, NGC~4372 has already been identified as a potential host for a large population of stellar mass BHs \citep{abbas2018, arcasedda2018, rui2021}.





\begin{figure*}
  \centering
  \includegraphics[width=\textwidth]{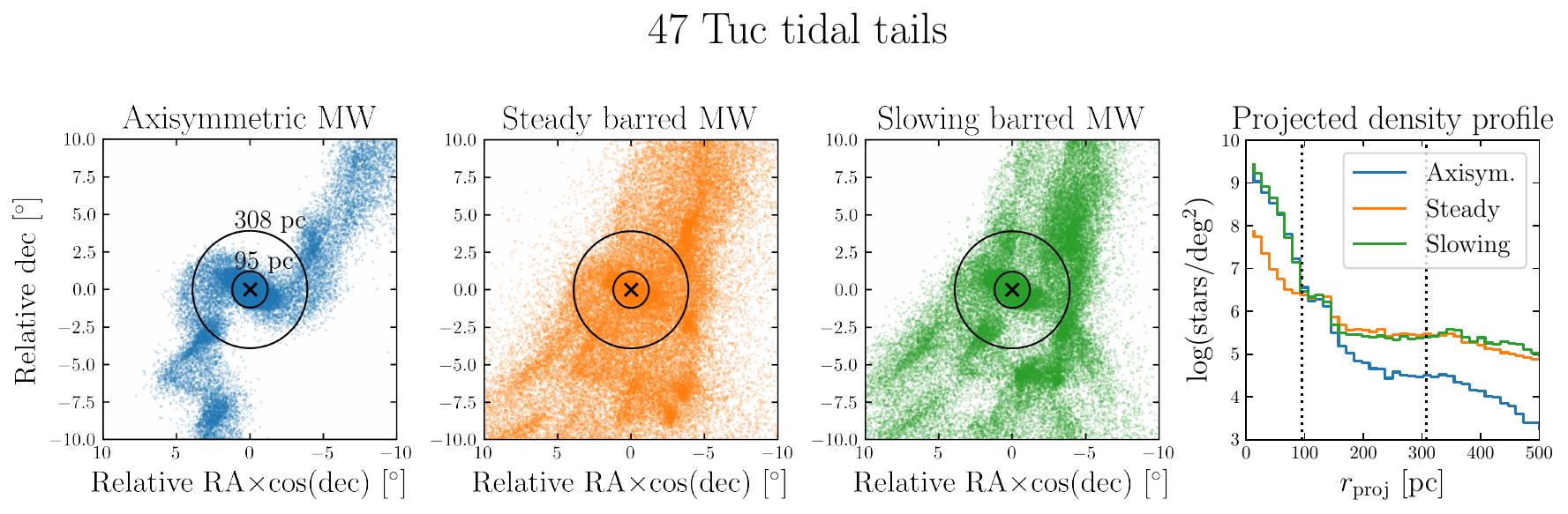}
  \caption{Final snapshots of the tidal tail simulations of 47 Tuc in the axisymmetric (left-hand panel), steadily rotating barred (second panel) and slowing barred (third panel) Milky Way potentials. The particles are shown in on-sky coordinates centred on the cluster, marked by the black cross. The two black circles indicate the radial range (95-308 pc) of the diffuse extended halo observed by \citet{piatti2017}. The right-hand panel shows the projected density of the stripped stars as a function of projected radius, with the two radii again marked. The bar causes the tidal tails to be much more diffuse, creating an extended halo around the cluster with a flat density profile.}
  \label{fig:stream}
\end{figure*}

\section{Tidal tails of 47 Tuc}\label{section:stream}

We have found that 47 Tuc is the cluster most likely to be trapped in the corotation resonance based on uncertainties in its current position. We now consider the implications of this trapping on the structure of tidal tails composed of stars stripped from the cluster. Tidal tails (i.e. stellar streams) have been predicted to be present near 47 Tuc \citep{lane2012}, but \citet{piatti2017} have found no evidence of their presence, only a diffuse halo extending several times beyond the cluster's tidal radius. This is present on all sides of the cluster and has a nearly flat density profile between projected radii of $r_\mathrm{proj}\sim95$ pc and $\sim308$ pc. We run test particle simulations of tails formed from 47 Tuc to investigate whether these observations can be explained by the cluster being on a trapped orbit. In this section we neglect the slowing of the bar, and only consider the effect of a trapped orbit at constant pattern speed.

\subsection{Stream generation}
We generate stellar streams with a method similar to that described by \citet{dillamore2022}, using the Modified Lagrange Cloud Stripping technique \citep[e.g.,][]{gibbons2014,bowden2015}. This involves integrating the orbits of test particles in the combined potential of the Milky Way and the progenitor cluster, modelled as a Plummer sphere with mass $M_\mathrm{c}$ and scale radius $a_\mathrm{c}$. The particles are released from the $L_1$ and $L_2$ Lagrange points, situated at a distance $r_\mathrm{t}$ from the centre of the cluster along the line from the Galactic centre. This is defined by
\begin{equation}
    r_\mathrm{t}\equiv\left(\frac{GM_\mathrm{c}}{\Omega^2-\frac{\partial^2\Phi}{\partial r^2}}\right)^{1/3},
\end{equation}
where $\Omega$ is the instantaneous angular speed of the cluster about the Galactic centre and $\Phi$ is the Milky Way potential. The particles' velocity components are each drawn from a Gaussian distribution with velocity dispersion $\sigma$.

We set the mass and Plummer scale radius of 47 Tuc to $M_\mathrm{c}=7.79\times10^5M_\odot$ \citep{baumgardt2018} and $a_\mathrm{c}=4.32$~pc. The latter is such that the Plummer half-mass radius matches that reported by \citet{baumgardt2018}. The velocity dispersion of released particles is $\sigma=1$~km\,s$^{-1}$, following one of the models of \citet{lane2012}.

We use three different potentials: the axisymmetrised version of the \citet{sormani2022} Milky Way potential with no bar; the \citet{sormani2022} barred potential rotating at constant pattern speed $\Omegab=35$~km\,s$^{-1}$\,kpc$^{-1}$, so the orbit of 47 Tuc is identical to that shown in Fig.~\ref{fig:orbits}; and the decelerating barred potential described in Section~\ref{section:slowing_potential}. We first integrate the orbit of 47 Tuc back in time to $t=-2$~Gyr, then run the simulation forward to the present day while releasing the particles at the Lagrange points.

\subsection{Results}
The tidal tails produced by the simulations are shown in on-sky coordinates in Fig.~\ref{fig:stream} for the axisymmetric (left-hand panel) and barred (middle panels) Milky Way potentials. In the former case, two clear tidal tails are formed, similar to those predicted by \citet{lane2012}. However, in both barred potentials the stripped stars form a much more extended structure which surrounds the cluster on all sides. The two black circles mark the projected radii (95 and 308 pc) between which \citet{piatti2017} observed an extended diffuse halo around 47 Tuc. We see that in the barred Milky Way potentials this area is filled by the stripped stars on all sides of the cluster, producing a diffuse halo. The results for the steady and slowing barred potentials are similar, except that the distribution of stripped stars is more clumpy with the decelerating bar. The right hand panel shows the projected surface density of stripped stars as a function of projected radius in the three simulations. The black dotted lines mark 95 and 308 pc. The simulations in the barred potentials produce a much flatter surface density profile between these two radii than the axisymmetric case, consistent with the observations of a flat density profile by \citet{piatti2017}.

We therefore propose that the lack of observed tidal tails around 47 Tuc is due its orbit being trapped in the bar's corotation resonance. This explains the presence of a diffuse extended halo around the cluster.

The morphology of the tidal tails in the barred potential results from the libration of the cluster around the resonance \citep[see pp.193-196 of][]{binney_tremaine}. Unlike in the axisymmetric case, the orbital frequencies of the cluster and its stripped stars oscillate, allowing some stars to return to the vicinity of the cluster some time after being stripped. The extended diffuse halo around the cluster comprises these stars.

\section{Conclusions}\label{section:summary}
We have considered the influence of the Milky Way's bar on the dynamics of globular clusters (GCs) via its resonances. Our principal findings are summarised below.

\begin{enumerate}[label=\textbf{(\roman*)}]
    \item A bar with a decelerating pattern speed is capable of causing large changes in the angular momenta $L_z$ and energies $E$ of GCs. By directly integrating the orbits of 141 GCs, we select 10 with the largest positive changes in $L_z$ due to a decelerating bar. 
    \item The selected GCs experience dramatic changes in their orbits due to the slowing bar. As the bar slows, they are transported to higher $E$ and $L_z$. In most cases the orbits are more eccentric and have smaller pericentres at $t\sim-5$~Gyr. In a few cases (e.g. M22) the orbits flip from retrograde to prograde as the bar slows.
    \item Six of these 10 GCs can be associated with the corotation resonance (CR) of the bar, including M22, 47 Tuc, and Pal 11. We call these \textit{trojan globular clusters}. These six clusters are all considered to have been born \textit{in situ} and are generally on thick disc-like orbits. Their phase space positions coincide with an overdensity of stars previously associated with the Hercules stream.
    \item Two of the GCs (M22 and M2) are candidate nuclear star clusters due to their spreads in metallicity. The bar may have transported M22 from much smaller Galactic radii ($r_\mathrm{peri}\approx0$) to its present position. This supports the theory that M22 could have been a nuclear star cluster of the primordial Milky Way.
    \item We simulate tidal tails of the trojan GC 47 Tuc in Milky Way potentials with and without a rotating bar. In the axisymmetric potential a pair of narrow tails is formed from the cluster. However, trapping the cluster around the CR of a barred potential causes the tails to become much more blurred. An extended diffuse halo of stripped stars with a flat projected density profile is formed around the cluster, consistent with observations of 47 Tuc \citep{piatti2017}.
\end{enumerate}

Globular cluster orbits may also be affected by a variety of other factors not considered in this study, such as a time-dependent potential or changing disc orientation \citep[e.g.][]{dillamore2022,nibauer23}. Recent accretion of satellites is less likely to be important, since there have been no significant merger events in the inner Galaxy since the formation of the bar \citep{sanders2023,deason24}. Dynamical friction (DF) may also act to decrease the energy of clusters, counteracting resonant dragging. However, the cluster in our sample most affected by DF \citep[NGC 5927;][]{moreno22} experiences a change in energy of $\dot{E}\sim-2\times10^2$~km$^2$\,s$^{-2}$\,Gyr$^{-1}$, only $\sim5\%$ of that due to the bar. Other trapped GCs are even less affected, so DF is unlikely to significantly affect their orbits. However, these time-dependent effects serve as a reminder that the globular cluster system of the Milky Way is not in steady state, and `integrals of motions' are not generally conserved. Future Galactic archaeology studies using GCs should take resonances and other perturbations into consideration.

\section*{Acknowledgements}
We thank Holger Baumgardt and the Cambridge Streams group for helpful comments and suggestions during this study. We are grateful to the anonymous referee whose comments have helped to improve this manuscript. AMD thanks the Science and Technology Facilities Council (STFC) for a PhD studentship. VB, SM and NWE acknowledge support from the Leverhulme Research Project Grant RPG-2021-205: `The Faint Universe Made Visible with Machine Learning'.

This work has made use of data from the European Space Agency (ESA) mission
{\it Gaia} (\url{https://www.cosmos.esa.int/gaia}), processed by the {\it Gaia} Data Processing and Analysis Consortium (DPAC,
\url{https://www.cosmos.esa.int/web/gaia/dpac/consortium}). Funding for the DPAC has been provided by national institutions, in particular the institutions participating in the {\it Gaia} Multilateral Agreement.

This research made use of Astropy,\footnote{http://www.astropy.org} a community-developed core Python package for Astronomy \citep{astropy:2013, astropy:2018}. This work was funded by UKRI grant 2604986. For the purpose of open access, the author has applied a Creative Commons Attribution (CC BY) licence to any Author Accepted Manuscript version arising.

%

\vspace{5mm}







\bibliography{refs}{}
\bibliographystyle{aasjournal}



\end{document}